# Polaronic effect and its impact on $T_c$ for novel layered superconducting systems.

Vladimir Kresin

Lawrence Berkeley National Laboratory,

University of California, Berkeley, CA 94720

Abstract

The crystal lattice of a complex compound may contain a subsystem of ions with each one possessing two close equilibrium positions (double-well structure). For example, the oxygen ions in the cuprates form such a subsystem. In such a situation it is impossible to separate electronic and local vibrational motions. This leads to a large increase in the effective strength of the electron-lattice interaction, which is beneficial for pairing.

## 1. Introduction

This paper is concerned with the lattice dynamics and the impact of electron-lattice interaction on the properties of novel superconductors, especially the high $T_c$ cuprates. More specifically, we address the question: why the presence of polaronis states is beneficial for superconductivity?. As is known, polaronic effect provided the main motivation for the original search for high $T_c$ in the cuprates [1].

Consider a complex compound where one ionic sub-system is characterized by two close equilibrium positions (double-well potential). Such strongly anharmonic potential leads to a peculiar non-adiabatic polaronic effect; indeed, in this case the electronic and local lattice degrees of freedom turn out to be un-separable (see below). In fact, oxygen ions in the cuprates do form such a sub-system; the double-well structure has been observed experimentally in [2] and studied theoretically by S.Wolf and the author in [3], see also the reviews [4,5]. One can demonstrate – and this is the goal of the present Letter - that the presence of such a structure leads to a noticeable increase in the effective strength of electron-lattice coupling relative to the usual case of a single potential minimum for the ionic coordinate.

An impact of non-linear lattice dynamics, including anharmonicity, on the pairing has been studied in a number of interesting papers (see,e.g., [6,7]); direct interaction of electrons and non-harmonic lattice has been studied. Here our focus is different. The thing is that the pairing interaction is described by the matrix element connecting an initial and virtual states. One can show that even in the absence of direct interaction with the double-well structure, the resulting coupling constant will increase . This is due to the increased phase space for virtial transitions or, in other words, due to an increased number of these transitions. Below we describe the corresponding formalism.

## 2. Double-well structure; diabatic representation

According to the adiabatic theory, the equation

$$\hat{H}_{\vec{r}}\psi_n(\vec{r},\vec{R}) = \varepsilon_n(\vec{R})\psi_n(\vec{r},\vec{R}); \hat{H}_{\vec{r}} = \hat{T}_{\vec{r}} + V(\vec{r},\vec{R}) \qquad (1)$$

determines the electronic terms $\varepsilon_n(\vec{R})$ which depend parametrically on ionic positions. Here $\{\vec{r},\vec{R}\}$ are the electronic and ionic coordinates, $\hat{T}_{\vec{r}}$ is the operator of kinetic energy of electrons, $V(\vec{r},\vec{R})$ is the total potential energy, and $\psi(\vec{r},\vec{R})$ is the electronic wave function. The electronic term forms the potential for the ionic motion.

Consider now the special case when the electronic term for some ionic subsystem (e.g., for oxygen ions in the cuprates) contains two close minima positions. As an example, we focus on the case when the double-well potential corresponds to some direction ; for layered systems this direction is perpendicular to the layers (OZ||c). As for the dependence of the ionic motion on X,Y, it is described by an usual harmonic dynamics with single minima equilibrium positions. Therefore, $\varepsilon_n(\vec{R}) \equiv \varepsilon_n(\vec{\rho},Z)$ with the double-well structure in the Z direction, where $\bar{\rho}$ denotes the in-plane ionic position.

We employ the tight-binding approximation. As a first step, we should write down the local wave function. In our case, the ion is affected by the double-well potential (ine Z direction; Fig.1). At this stage is very convenient to employ the diabatic representation [8-10]. As a result of the transformation, the double-well potential is replaced by two crossing harmonic energy terms (Fig.1). Each of these terms contains also its vibrational manifold. Therefore, the local state is described by two groups of terms ("a" and "b"), so that

$$\Psi_{loc.} = c_a \Psi_a(\vec{r},\vec{R}) + c_b \Psi_b(\vec{r},\vec{R})$$
$$where$$
$$\Psi_{a(b)}(\vec{r},\vec{R}) = \psi_{a(b)}(\vec{r},\vec{R}_0^{a(b)}) \chi_{a(b)}(z - z_0^{a(b)})$$

(2)

Note that $\Psi_a = \psi_s \chi_{sv}; \Psi_b = \psi_{s'} \chi_{s'v'}$. Here {s,s'} and {v, v'} are the electronic and vibrational quantum numbers. If both terms are equivalent, then $c_a^2 = c_b^2$ =0.5. Because of the inter-term tunneling, the energy levels which correspond to isolated terms ,are splitted into symmetric and antisymmetric ; the scale of the splitting is of order of $\varepsilon_{ab}$, where (see [10])

$$\varepsilon_{ab} = \int d\vec{R}\chi_a \hat{H}_{\vec{r};ab}\chi_b$$
$$\hat{H}_{\vec{r};ab} = \int d\vec{r}\psi_b^* \hat{H}_{\vec{r}}\psi_a \quad (3)$$

It is essential that, contrary to the usual adiabatic picture, the operator $\hat{H}_{\vec{r}}$ defined by Eq.(1) has non-diagonal terms in the diabatic representation.

One can see directly from Eq.(2) that the wave function $\Psi_{loc.}$ can not be written as a product of electronic and ionic wave functions . In other words, the electronic and local ionic motions can not be separated like in the usual adiabatic theory. Such a polaronic effect is a strong non-adiabatic phenomenon .

In the tight-binding approximation each local term, including those formed by the transformation (Fig.1) is broadened into the energy band. Note at first that the local electronic function (see Eq.(2)) $\psi_i(\vec{r},\vec{R}_0^i)$ (i=a.b) can be written in the form: $\psi_i(\vec{r}-\vec{R}_0^n), \vec{R}_0^n$ corresponds to the crossing point for n-th ion (i=a,b). Indeed, because the electronic wave function has a scale of the length of the bond that greatly exceeds the amplitude of vibrations and the distance "δ" between the minima, one can neglect the difference between $\vec{R}_0^n$ and $\vec{R}_0^i$.

The total wave function contains also the vibrational part. At first, let us separate the local vibrational mode that corresponds initially to the double-well structure, or, in the diabatic approximation, to two crossing terms (see above). As for the dependence of ionic motion on X and Y, it can be described by a set of normal modes. Then the total wave function has a form:

$$\Psi(\vec{r},\vec{R}) = u_{\vec{k}}(\vec{r},\vec{R})e^{i\vec{k}\vec{r}}\Phi_{vib.}$$
$$u_{\vec{k}}(\vec{r},\vec{R}) = \sum_n \Psi_{loc.}(\vec{r},\vec{R})e^{i\vec{k}(\vec{R}_0^n - \vec{r})} \quad (4)$$

$\Psi_{loc.}$ is determined by Eq.(2). The local vibrational wave functions are centered at $z_0^i$ and at $z_0^i - \delta$. If the terms are similar, then $\chi^b(z) = \chi^a(z-\delta)$.

One should stress a key difference between the ionic motions described by the function

$$\Phi_{vib.}(R_x, R_y) = \Pi\varphi(\Omega_m) \quad (5)$$

($\Omega_m$ are the frequencies of the usual normal modes) and the function $\chi(R_z)$. There is no the in-plane ionic hopping between various ionic sites. However, the oxygen can tunnel between the two minima, and this is reflected in a noticeable overlap of vibrational functions $\chi^i$ corresponding to the two crossing terms (Fig.1).

## 3. Superconducting state

Let us focus on the electron-lattice interaction and its impact on pairing. We consider the case when the main contribution to the interaction is coming from the interaction with usual normal (harmonic) modes $Q_m$ (one can easily to describe a more general case). Then the interaction Hamiltonian can be written in the usual form:

$$\hat{H}' = (\partial V/\partial Q_m)\delta Q_m \qquad (6)$$

Summation over m is implied. The equation for the pairing amplitude has a form:

$$\Delta(\omega_n) = T \sum_{\omega_{n'}} \int \varsigma d\xi N_F D(\omega_n - \omega_{n'}; \tilde{\Omega}) F^+(\omega_{n'}) \qquad (7)$$

Here $D = \tilde{\Omega}^2[\tilde{\Omega}^2 + (\omega_n - \omega_{n'})^2]^{-1}$ is the phonon propagator, $\omega_n = (2n+1)\pi T$, $\tilde{\Omega}$ is the characteristic phonon frequency, $F^+(\omega_n) = \Delta(\omega_n)[\omega_n^2 + \xi_{m\nu}^2 + \Delta^2(\omega_n)]^{-1}$ is the Gor'kov's pairing function [11], see, e.g., [12], $\xi_{n\nu}$ is the excitation energy relative to the Fermi energy, $N_F$ is the density of states). It is essential that $\xi_{n_v}$ is the energy of virtual transitions which contribute to pairing. Note that the integrand contains the matrix element ($\zeta = g^2$) which has a form

$$g = \int \Psi^{*f} \hat{H}' \Psi^i d\vec{r} d\vec{R} \qquad (8)$$

where $\hat{H}'$ is defined by Eq.(6), i and f denote the initial and virtual states, and the total function has a form (4).

Let us introduce some characteristic frequency $\tilde{\Omega}$ and a single mode $\tilde{Q}$ which provides a main contribution to the pairing (cf.Eq.(7)). Then the product $\Pi \varphi_{v_m}(Q_m)$ can be replaced by the function $\varphi_v(\tilde{Q})$. Correspondingly H'= $(\partial V/\partial \tilde{Q})_0 \delta \tilde{Q}$. In other words, we model the phonon spectrum as consisting of two modes. One of them, harmonic mode $\tilde{Q}$, provides a major contribution to the coupling interaction, and the second mode which is characterized by double well structure (Fig.1), provides an additional number of the virtual states. This model can be easily generalized. One can see from Eq.(8) that the average value of the coupling matrix element is a sum:

$$g_{av.} = g_0 + g_1 \qquad (9)$$

where

$$g_0 = \int d\vec{r} d\tilde{Q} \sum_n \psi_s(\vec{r} - \vec{R}_0^n) \psi_s(\vec{r} - \vec{R}_0^n) \left(\partial V / \partial \tilde{Q}\right)_0 \delta\tilde{Q}_{lav.}$$

$$g_1 \approx g_0 \overline{F}^{ab} \qquad (9')$$

where

$$F^{ab} = 2 \sum_{s,v_1} \int \chi_{sv}(z - z_0^a) \chi_{s_1 v'}(z - z_0^b)$$

(9")

As usual (see.e.g., [13]), the displacement $\delta\tilde{Q}$ can be expanded in the Fouirie series.

It is very essential that the Franck-Condon factor $F^{ab} \neq 0$. Indeed, Eq.(7) assumes summation over virtual states, that is integration over $\xi'$ and summation over $v'$. The vibrational wave functions for the same term are orthogonal. However, this is not the case for different crossing terms.

The additional contribution $g_1$ corresponds to virtual transitions which are accompanied by the change of electronic terms upon ionic tunneling between the two minima. In other words, $g_1$ corresponds to transitions to the states with various $v'$ (for given $v$, see Eq.(2)). It is important to note that the contribution comes only from the terms with the same symmetry, that is from the terms with the same sign of the coefficient "b" (see Eq.(2)).

Therefore, there is an additional contribution (relative to the usual case), caused by the transitions between the local vibrational levels. Such an increase in phase space for virtual transitions leads to an increase in the value of the coupling matrix element g.

Consider in more detail the Franck-Condon factor $F^{ab}$. Assume for concreteness that the initial state corresponds to $v_a=0$. Eq. (7) contains a sum $\sum_{v'} F^{ab}_{v'0}$ which represents the contributions of virtual transitions to the manifold of the coherence states. In the diabatic representation (see above), a single non-harmonic term is replaced by two crossing terms, and each of them can be treated in harmonic approximation. The Franck-Condon factor is equal (see, e.g., [14]): $F^{ab} = (\bar{\beta}^v / v!) \exp(-\beta)$, where $\beta = (\delta/\alpha)^2$ ($\alpha$ is the amplitude of vibrations). One can see that $F_{v'0}$ contains a small factor: $\exp[-(\delta/\alpha)^2]$ which describes the probability of the ionic tunneling between two minima. But this smallness is compensated by summation over all virtual transitions. If, for example, $(\delta/\alpha)=2$, then $g_1 \approx 0.7 \, g_0$. If $g_0 \approx 1$, when g= $g_0$+ $g_1 \approx 1.7$. The value of $T_c$ is determined by the coupling constant $\lambda \propto \zeta \propto g^2$, and we obtain $\lambda \approx 3$. Therefore, the polaronic effect leads to a large increase in the effective strength of the electron-lattice coupling. As was stressed above, this enhancement arises due to the increase in phase space for virtual transitions which promote the pairing.

In summary, the electron-lattice interaction was studied for the case of a complex system when the electronic term (potential energy surface) for some ionic degree of freedom contains two close equilibrium positions (double-well structure). An example of such a system is offered by the oxygen ions in cuprates. Using the diabatic representation, one can describe this anharmonic case as motion in the potential formed by two crossing harmonic terms. The presence of such terms results in the formation of coherent ionic states with a concomitant increase in the number of virtual transitions. Importantly, these very transitions are the key ingredient entering the equation for the pairing order parameter. Since the vibrational wave functions for different terms are not orthogonal, the inter-term transitions are not forbidden. The increased phase space for virtual transitions brings about an increase in the coupling constant and thereby in the value of $T_c$.

The author is grateful to L.Gor'kov for interesting discussion. The research was supported by DARPA.

**References**

1. G. Bednorz, and K. Mueller,1986. Zeitshrift fur Physik B 64,189
2. D. Haskel , E. Stern , D.Hinks , D. Mitchell , J.Jorgenson , Phys.Rev.B56, 521 (1997).
3. V.Kresin and S.Wolf, Phys.Rev.B49, 3652 (1994)
4. V.Kresin, Y.Ovchinnikov and S.Wolf, Physics Reports, 431,231 (2006);
5. V.Kresin and S.Wolf, Rev. of Mod. Phys. 81,481(2009)
6. J . Hui and P.Allen,J.Phys.F 4,L42 (1974); J.Hardy and J.Flocken, Phys.Rev.Lett. 60,2191 (1988)
7. A, Bussmann-Holder and A,Bishop, Phys.Rev.B56,5297 (1997); J.of Super.&Novel Magnetism 10,1557 (1997)
8. T. O'Malley, Phys. Rev.252,98 (1967)

9.10. V.Kresin, J.of Chem.Phys. 128, 094706 (2008)

11. L. Gor'kov , JETP 7, 505 (1958.);

12. A.Abrikosov, L. Gor'kov, I. Dzyaloshinski, *Methods of Quantum Field Theory in Statistical Physics*, Dover, NY (1963)

13. A.Anselm, *Introduction to Semiconductor Theory*, Prentice Hall, New York (1982)

14. L.Landau and E.Lifshits, *Quantum Mechanics*, Sec.41, Pergamon Press,New York,NY (1976)

**Figure Caption:**

Fig.1. Transformation to the diabatic representation. Solid line: initial term (double-well structure); broken lines: crossing terms

## Crossing of Terms

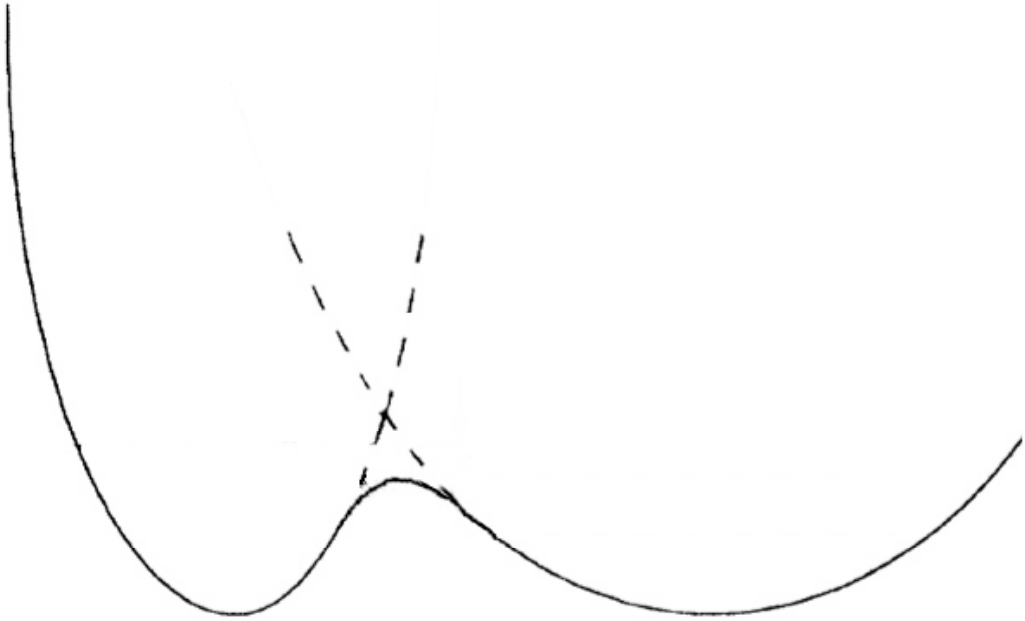